\documentstyle[newpasp,epsfig,psfig]{article}  
\def\simlt{\lower.5ex\hbox{\ltsima}}   
\def\gtsima{$\; \buildrel > \over \sim \;$}
\def\simgt{\lower.5ex\hbox{\gtsima}}

\begin{document}

\title{Simulations of unstable gaseous disks and the origin of giant planets}

\author{Lucio Mayer, Thomas Quinn}
\affil{Department of Astronomy, University of Washington, Stevens Way, 
Seattle, WA 98195 and NASA Astrobiology Institute}
\author{James Wadsley}
\affil{Department of Physics \& Astronomy, McMaster University, 1280 Main St. West, Hamilton ON L8S 4M1 Canada}
\author{Joachim Stadel}
\affil{University of Victoria, Department of Physics and Astronomy, 
3800 Finnerty Road, Elliot Building, Victoria, BC V8W 3PG Canada}

\begin{abstract}

We study the evolution of cold, gravitationally unstable protoplanetary gaseous disks performing 3D SPH simulations with up to a million particles on large parallel machines. We show that self-gravitating protoplanets can form in disks 
with initial $Q_{min} \sim 1.4$ provided that disks are evolved with a locally isothermal equation of state until localized regions reach a sufficiently high density. Lower resolution simulations carried out for as long as a thousand years show that systems of giant protoplanets with masses and orbits comparable to those of the observed extrasolar planets arise naturally.

\end{abstract}

\section{Introduction}

In recent years a large number of extrasolar planets has been discovered
via the wobble they induce on their star (Marcy \& Butler 1998).
Such planets have masses 
ranging from comparable to Jupiter to ten times larger and have orbits 
going from nearly circular to very eccentric (Marcy \& Butler 2000). 
When planets have been also
detected by means of photometric transits
their radius has been determined, indicating that they truly are gas giants
(Brown et al. 2001).
In the standard core-accretion
model giant planets might require longer than $10^6$ years to
form (Pollack et al. 1996) which would exceed observed disk
lifetimes (Haisch, Lada \& Lada 2001).
An alternative scenario, namely
the formation of giant planets by fragmentation of a gravitationally unstable 
gaseous protoplanetary disk, would
occur at least a thousand times faster (Boss 1998).  
Transient overdensities
have been observed in simulations of the disk instability, but whether they
can survive and become actual
protoplanets has been a puzzle so far (Boss 2001, Pickett et al. 2001). 
An extremely high resolution is needed to properly represent the self-gravity 
of the developing overdensities and to prevent them from being artificially sheared apart
by the stellar tidal force (Boss 1998,2001). 
In addition, restrictive boundary conditions
can enhance instabilities or suppress already formed overdensities that happen to move 
out of the computational volume (Pickett et al. 2001).
Smoothed particle hydrodynamics
(SPH) simulations describe the gaseous medium as a collection of particles 
and, being spatially adaptive by construction, can follow arbitrarily 
high densities.
Such simulations can thus properly follow the overdensities once a 
sufficiently large number of particles is employed and numerical issues 
like artificial fragmentation are overcome (see Nelson 1998 and these proceedings; Bate \& Burkert 1997).  Here we present new 3D SPH
simulations of unprecedented resolution showing for the
first time that long-lived, self-gravitating protoplanets arise as a result of
fragmentation in disks with masses comparable to those inferred for
the protosolar nebula.

\section{Initial Conditions and Simulations}

Our protoplanetary disk models closely follow those of Boss (2001) who used
a different numerical technique, namely a grid code with fixed boundaries.
We start from smooth, axisymmetric self-gravitating disks which extend 
from 4 to 20 AU and are in near-keplerian rotation around a solar mass star 
represented by a point mass. The stellar potential is exactly keplerian at 2.5 
AU and it is softened on smaller scales to speed up the computation.
The gravitational softening of the gas particles is chosen small enough 
to resolve gravity at least as well as the hydrodynamical forces up to
the onset of the nonlinear stage (the resolution of the hydrodynamical
forces is variable in SPH and increases as the local density increases).
The simulations were performed with GASOLINE, a highly parallel N-Body/SPH
code widely used in cosmology/galaxy formation.
Thanks to the multistepping capabilities of the code which allow to follow a huge 
dynamic range we adopted free boundaries.
The disks have a minimum Toomre Q parameter of either $1.4$ or $1.75$, 
reached at the outer disk boundary, and masses of, respectively, $0.1$ 
and $0.08 M_{\odot}$.

\begin{figure}
\plotone{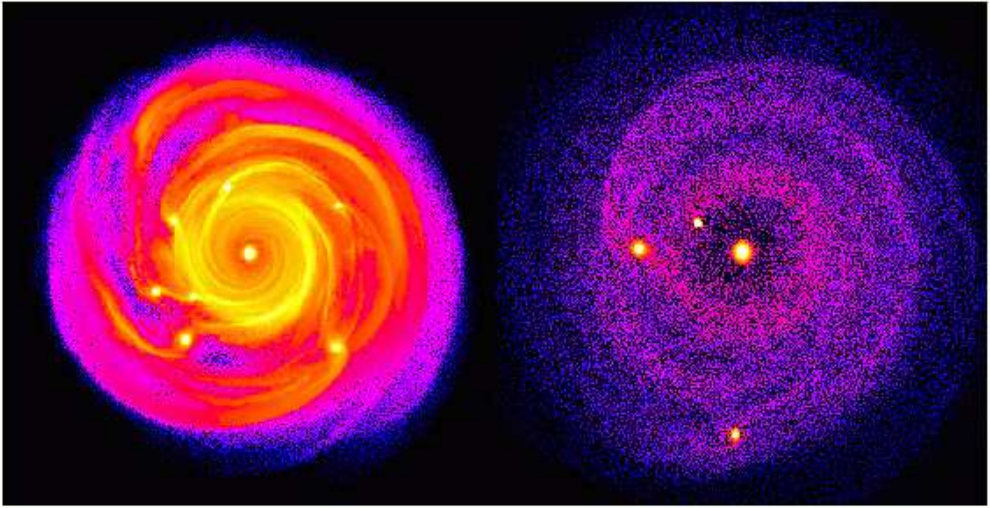}
\caption{Colour-coded logarithmic density plots of the disk with initial 
$Q_{min}=1.4$
seen face-on after 350 years in the hi-res simulation (left) and after 
850 years in the corresponding lo-res run (right). Density ranges between $10^{-14}$ and $10^{-7}$ g/cm$^3$ (brighter colors are for higher densities).
Boxes extend out to 25 AU. We show the runs evolved
adiabatically once overdense regions reach a threshold density (see text).}
\end{figure}

The initial temperature follows a power 
law profile (temperatures are $50$ K at $R > 10$ AU and reach $650$ K at 4 AU)
predicted by detailed calculations of thermal balance between 
the central star, the disk and the protostellar cloud (Boss 1998).
Recent models of protoplanetary disks yield outer disk
temperatures at least as low as those used here (D'Alessio, Calvet \& Hartmann  2001).
As for their thermodynamical evolution, disks are evolved either maintaining the
initial temperature profile (a {\it lagrangian} locally isothermal
approximation, i.e. the temperature of a given bucket of particles is 
kept constant) or 
allowing them to keep the heating resulting from adiabatic compression
and/or shocks 
once sufficiently strong overdensities appear. 
The second type of simulations accounts for the fact
that the gas would behave nearly adiabatically once
its mean density grows beyond a certain value (Boss 2001)- it
will not be able to radiate away its internal energy as efficiently as 
assumed with the locally isothermal approximation. 
We performed simulations with as many as 1 million gas particles as well
as lower resolution simulations. We found numerical results
to nearly converge with at least 200.000 particles; the latter faster
computations were used to study disks for longer timescales.

\section{Results}

After $\sim 150$ years (5 orbital times at the reference radius of 10 AU) 
all disks
develop trailing spiral arms and local overdensities at $R > 10$ AU.
While in disks with $Q \sim 1.75$ the spiral pattern grows and then reaches
a weaker, near-stationary pattern up to $500$ yr, in disks with $Q \sim 1.4$
fragmentation occurs along the spiral arms, and several clumps appear. Subsequently other clumps form down to $7-8$ AU (inside this radius the disk is too
and non-axisymmetric instabilities are suppressed). Once formed, these clumps contract 
reaching densities as high as $10^{5}$ times the initial local value on their free-fall timescale (tens of days); 
their collapse then is nearly halted as the interparticle separation becomes much smaller than the gravitational softening and the resolution limit of our calculation is reached - in this regime gravity is artificially weaker than pressure forces (Bate \& Burkert 1997).

\begin{figure}
\plotfiddle{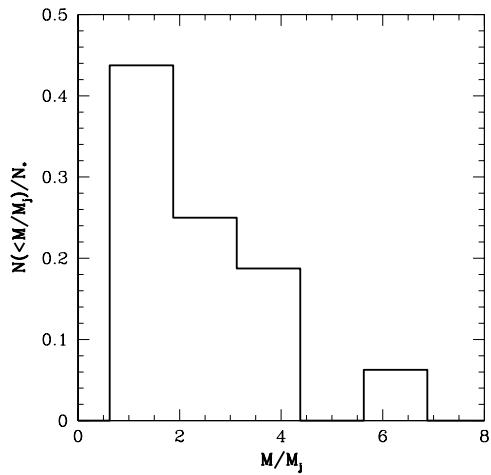}{5cm}{0}{35}{35}{-100}{-60}
\caption{Mass function of clumps after 350 years in the hi-res disk with initial $Q_{min}=1.4$. Material with densities at least ten times higher than the initial local density is associated with the clumps.}
\end{figure}

These highly dense clumps are self-gravitating and easily resist stellar tides (Figure 1).
Clump formation and collapse is basically independent on how the disk evolves thermodynamically once regions with densities at least ten times higher than the initial local density have appeared. 
Individual clumps can collide and merge with other clumps. As shown in Figure 2,after 350 years protoplanets exhibit a mass function qualitatively close to that of observed extrasolar planets (e.g. Marcy \& Butler 2000) - planets smaller than Jupiter likely do not form because of resolution limits. Once clumps collapse to very high densities it becomes increasingly difficult to follow the disk evolution as computational timesteps becomes very small. We thus resort to the simulations with 200.000 particles (which also have particle softening increased by a factor of three for numerical consistency - see Dehnen 2002). In these runs a smaller number of clumps  forms
due to the lower resolution but their dynamical evolution is followed for up to nearly 10$^3$ years. At the end of the simulations only three protoplanets remain with masses between 2 and 6 $M_{Jup}$ and these are on orbits ranging from nearly circular to an eccentricity of about 0.3, between 3 and 15 AU (Figure 1).
This final stage is reached several orbital times before the simulations are ended and therefore we believe it is fairly representative of the dynamical evolution of protoplanets formed
in this instability model. Of course, both orbital migration (Bryden et al. 1999) and secular dynamical interactions between the protoplanets (Terquem \& Papaloizou 2002) can still occur on timescales much longer than those explored here and might change the orbits.

\section{Conclusions}

Our SPH simulations answer for the first time one of the fundamental question regarding the instability model; can overdensities survive and turn into tidally stable, self-gravitating protoplanets? We find that they can once a sufficiently high resolution is used and no restrictions are imposed on the dynamical evolution of the disk. Nearly all the clumps at some point in their evolution reach distances from the central star that are either larger or smaller than the initial disk boundaries and would have been lost in a simulation that preserves the latter. We also show that systems of giant planets with masses and orbital eccentricities close to those of observed extrasolar planets naturally arise. Therefore, we conclude that the gravitational instability picture is a viable model for giant planet formation.
Whether keeping the
disk locally isothermal until the first strong overdensities appear is physically motivated  needs to be explored in more detail.
For this to happen radiative cooling must be very efficient at dissipating the heat resulting from compressions and shocks occurring during the
earlier stages (Pickett et al. 2001). In addition, future work will also be directed 
to find a more realistic setup of the initial conditions using a disk actually 
grown from the protostellar cloud. This will allow the investigation of how
likely is a situation in which a disk has $Q_{min} \sim 1.4$; indeed mild spiral instabilities at an earlier stage, when the disk is lighter or hotter,
could drive it rapidly towards a much higher $Q$ (Laughlin \& Roczyska 1994).
However, the model with $Q_{min} \sim 1.75$ still has a $Q \sim 1.7$ after 350 years, suggesting that weak spiral instabilities are not very efficient in redistributing mass and stabilizing the disk.

\end{document}